# Simulating Raga Notes with a Markov Chain of Order 1-2

Devashish Gosain, Deptt. of Computer Sc & Engg, BIT Mesra, Ranchi-835215, India

Soubhik Chakraborty, Deptt. of Mathematics, BIT Mesra, Ranchi-835215, India

Mohit Sajwan, Deptt. of Computer Sc & Engg, BIT Mesra, Ranchi-835215, India

Email: gosain.devashish@gmail.com

soubhikc@yahoo.co.in

mohit.sajwan@gmail.com

**Abstract:** Semi Natural Algorithmic composition (SNCA) is the technique of using algorithms to create music note sequences in computer with the understanding that how to render them would be decided by the composer. In our approach we are proposing an SNCA2 algorithm (extension of SNCA algorithm [1]) with an illustrative example in Raga Bageshree. For this, Transition probability matrix (tpm) was created for the note sequences of Raga Bageshree, then first order Markov chain (using SNCA) and second order Markov chain (using SNCA2) simulations were performed for generating arbitrary sequences of notes of Raga *Bageshree*. The choice between first and second order Markov model, is best left to the composer who has to decide how to render these music notes sequences. We have confirmed that Markov chain of order of three and above are not promising, as the tpm’s of these become sparse matrices.

## 1. Introduction:

The term algorithmic composition refers to the science and art of creating music by using an algorithm (the term algorithm stands for a finite set of instructions for solving a problem, named after al-Khwarizmi, the famous Persian mathematician). Algorithmic composition is clearly a science, as it is a body of systematized knowledge that can be rationally learned and taught. It is also an art, in that it is a skill that can only be developed by continuous practice [1].

H.F Olson first used markov process in statistical analysis of music composition around 1950 based on 11 melodies of Stephen Foster. Then in 1956, Hiller and Isaacson of University of Illinois used markov models of variable order to produce first computer generated compositions [2, 3]. But the pioneering use of the computer in algorithmic composition is that of Iannis Xenakis, who created a program that would produce data for his "stochastic" compositions, which he had written about in great detail in his book Formalized Music (1963). Xenakis used the computer's high-speed computations to calculate various probability theories to aid in compositions like Atrées (1962) and Morsima-Amorsima (1962). The program would "deduce" a score from a "list of note densities and probabilistic weights supplied by the programmer, leaving specific decisions to a random number generator" [3, 4]. "Stochastic" is a term from mathematics which designates such a process, "in which a sequence of values is drawn from a corresponding sequence of jointly distributed random variables" (Webster's dictionary).

With Xenakis, it should be noted, however, "the computer has not actually produced the resultant sound; it has only aided the composer by virtue of its high-speed computations" [2, 4]: in essence, what the computer was outputting was not the composition itself but material with which Xenakis could compose [2].

SNCA2 algorithm also relies on the same theory. We believe that purely computer based algorithmic composition is more suited to music with fixed scores, such as Western music, than extempore music, such as Indian classical music because emotion and devotion are of paramount importance in any raga rendition and neither can be produced by calculated artistry. In present scenario approaches like fractals, artificial neural networks, Bayesian networks etc. are being used by many computer scientists and music researchers for artificial compositions. However some musicians do not approve artificial composition as it destroys musical creativity. Striking a balance between two Chakraborty et al. (2011) introduced Semi-Natural Composition where a computer generates only the sequence of music notes using first order markov process and the rest is decided by the composer. The present work is its extension in which we have experimented with both first and second order markov process with encouraging results.

***Note:*** Indian classical music has two domains-North Indian or Hindustani music and South Indian or Carnatic music. In either form, the nucleus is a raga. A raga is a melodic structure with fixed notes and a set of rules characterizing a certain mood conveyed through performance. By "a set of rules" we mean the typical way the permissible musical notes are to be used in the raga, the typical note combinations, and the aroh-awaroh (ascent-descent) pattern to be followed and so on. These rules help in characterizing the raga mood. *However, the way the notes are approached and rendered in musical phrases and the mood they convey are more important in defining a raga than the notes themselves*. In the North Indian musical tradition, ragas are associated with different times of the day, or with seasons (e.g.Bageshree is a North Indian (Hindustani) classical raga of the late night). Readers interested to know more about Indian classical music, especially Hindustani form, and its comparative similarities and differences with Western classical music are referred to chapter one of the book by Chakraborty, Mazzola, Tewari and Patra (2014).

## 2. Motivation

### Music has strong therapeutic effects on human body and psych

Music therapy, *when used with conventional treatment* can help to reduce pain and relieve chemotherapy-induced nausea and vomiting. It may also relieve stress and provide an overall sense of well-being. Some studies have found that music therapy can lower heart rate, blood pressure, breathing rate and can reduce labor pain in women [5-10]. According to the American Music Therapy Association, "Music Therapy is the clinical and evidence-based use of music interventions to accomplish individualized goals within a therapeutic relationship by a credentialed professional who has completed an approved music therapy program."

Singh et al. (2012) also reveals that recovery of patients (with Cognitive Impairment in the Cerebrovascular Accident and Head Injuries) who have been exposed to raga therapy recovers faster than others who have been deprived of it [11-14].

SNCA [I] and SNCA2 algorithms can help the music therapist to produce different compositions based on *Hindustani ragas/western music* which will lead to fast recovery of patients.

## 3. System Model:

In the present paper, initially a first order Markov chain [1] will be used to build an algorithm for generating arbitrary sequences of notes of raga Bageshree ( based on the SNCA Algorithm proposed by Chakraborty et. al. (2011) ). Thereafter a second order Markov chain will be used to build an algorithm for the previous kind. We call it as SNAC2 (an acronym for Semi-Natural Algorithmic Composition using 2$^{nd}$ order Markov Chain).

### 3.1. Markov Chain:

Markov chain represents a system of elements making transition from one state to another over time. The order of the chain gives the number of time steps in the past influencing the probability distribution of the present state, which can be greater than one. Markov Chains are stochastic processes that can be parameterized by empirically estimating transition probabilities between discrete states in the observed systems [15]. By a first order Markov chain, we mean a stochastic process where the future state depends only on the present state and not on the past one. By second order markov chain we mean a stochastic process where the future state depends not only on the present state but also on the past one. A stochastic process is a family of random variables that are functions of time measure, i.e., time need not be time of clock but an instance of realization of the response.

Let $X(t)$ be stochastic process, possessing discrete states space S = $(1, 2\ldots K)$. In general, for a given sequence of time points tl < t2 < ...< tn-1 < tn, the conditional probabilities should be:

$$Pr\ \{\ X(t_n)\ =\ i_n\ |\ X(t_1 = i_1\ \ldots.,X(t_{n-1}) = i_{n-1}\ \} = Pr\ \{\ X(t_n) = i_n\ |\ X(t_{n-1}) = i_{n-1}\ \}$$

The conditional probabilities $Pr\ \{X(t) = j\ |\ X(s) = i\} = P_{ij}(s, t)$ are called transition probabilities of order $r = t - s$ from state $i$ to state $j$ for all indices $0 \leq s < t$, with $1 \leq i, j \leq k$. They are denoted as the transition matrix $P$. For $K$ states, the first order transition matrix $P$ has a size of $K \times K$ and takes the form:

| | | | |
|---|---|---|---|
| p1,1 | p1,2 | ... | p1, k |
| p2,1 | p2,2 | ... | p2, k |
| ... | ... | ... | ... |
| pk, 1 | pk, 2 | ... | pk, k |

The state probabilities at time t can be estimated from the relative frequencies of the k states. A second order transition probability matrix can be shown symbolically as below

| | | | |
|---|---|---|---|
| p11,1 | p11,2 | ... | p11, k |
| p12,1 | p12,2 | ... | p12, k |
| ... | ... | ... | ... |
| p1k, 1 | p1k, 2 | ... | p1k, k |
| p21,1 | p21,2 | ... | p21, k |
| p22,1 | p22,2 | ... | p22, k |
| ... | ... | ... | ... |
| ... | ... | ... | ... |
| pkk, 1 | pkk, 2 | ... | pkk, k |

In this matrix the probability pijk is the probability of the next music note *k* if the current note is *j* and the previous note was *i.* This is how the probability of making a transition depends on the current state and on the preceding state. These matrices become the basis of future likely wind speed. The probability in any state varies between zero and one.
The summation of row in transition matrix is always equal to one. If the transition probability in the *ij row* at the *k state* is p*ijl*, then the cumulative probability is given by

$$P_{ijk} = p_{ij1} + p_{ij2} + p_{ij3} + .... + p_{ijk}$$

This cumulative probability helps in determining the future music note by using random number generator.

### 3.2. Generation of Transition Probability Matrix (tpm) for Second Order Markov Chain

Initially conditional probabilities are calculated. This is best explained with an example, say P(*n/DS*), the probability of next note to be n given that present note is *S* and previous note was *D*, is calculated by number of times *DS* is followed by *n*, divided by the total number of times *DS* pair occurs in the entire sequence. However, *DS* happened to be the last note pair in the *Bageshree* sequence, so there is no information of the next transition. Subtracting one from the number of times *DS* has occurred yields 10-1=9 possibilities of a transition. On 4 occasions, *DS* is followed by *n*. Hence we have that P(*n/DS*) = 4/9. Continuing the same procedure (49x7) size tpm is generated.

## 4. ALGORITHM SNCA2

Before examining the SNCA2 algorithm readers are suggested to refer to SNCA algorithm proposed by Chakraborty et. al. (2011) [1].

**Step 1**: Without any loss in generality, take the note at instance 1 to be the tonic *Sa*(*S*).
This is not mandatory; we can simulate the first note using the unconditional probabilities of the notes. The only danger is this can create a note which Indian music theory will not permit to be the starting note, e.g., Pa in Bageshree in our case which is a weak note in the raga. Sa can always be the first note in any raga. Also it is musically logical to start with the tonic, the base note or note of origin from which other notes are realized. Hence the precaution.

**Step 2**: Using the transition probabilities at *Sa*, one simulates the next note. The strategy is to generate a continuous uniform variate X in the range [0, 1] and, depending on the class in which X falls (see row one of Table II of first order markov giving the classes) the note at instance 2 is obtained as either S, r, g, M, P, d or n with respective transition probabilities as given in row 1 of Table I. The logic that defends the strategy is that since X is a continuous $U\,[0; 1]$ variate, so $P\,(a < X < b) = b\text{-}a$, obtained by integrating the probability density function of X, which is unity, between the limits a and b. Thus, for example, $P\,(13/46 < X < 22/46) = 9/46 = P\,(M/S)$ etc. [remark: the pdf of a continuous $U[r; s]$ variate is $1=(s\text{-}r)$].

**Step 3**: Using the pair of note simulated at instance 2 (present state), and note Sa (past state that was chosen initially) corresponding transition probability row of this note pair is selected (*in Table 3. of Second Order Markov Chain*), the next note at instance 3 is simulated similarly using cumulative probabilities to form classes.

**Step 4**: Using the notes generated at instance 3 (present state) and instance 2(past state) corresponding transition probability row of this note pair is selected(*in Table 3. of Second Order Markov Chain)*, the next note at instance 4 is simulated similarly using cumulative probabilities to form classes. The process is repeated.

## 5. Experimental Results

We first present the transition probability matrix of *Bageshree* using a first order Markov chain. Our results are given in Table 1. Next we provide the *Bageshree* class matrix in Table 2 using the entries in Table 1.
Extending this work, transition probability matrix of Bageshree is constructed using Second order markov chain. Results are depicted in table 4. Finally class matrix of Bageshree is created using the entries of table 3.

### Simulation of Raga

Applying algorithm SNCA[1] on the set of 1000 independent $U[0; 1]$ variates, using Sa (S) as the additional starting note, we immediately have a sequence of 1000 notes as our first illustrative example of raga *Bageshree* as given below:

**Example 1.**
nDMgRSMDSMDMPDSMDnDMDnDMPDMgRSRSSMgMDnDMDSnDMDnDMDMDMnD
MDnSRSDMDMn
nDnSnDRSMDMgRSRSMDMgMPDSnDMgMgRSnDSRSSDnDMDMDDMDSRSnSnDMPD
MgMDSSMDMgRSMSMgRSMDnDMnDMgMgRSnDnSnDSDSMDggMDSMgMSRSSnSRSn
DnDMDnSMDnDMgRSnDSSnDMnSnSnSnSnDnSnDMnDnDMMgRSnSMggMDMDgMDMg
MnDMnSSDMgMDMSnSnDMPDMDMgMnDMSnDMgRSnDMgRSnDDMDSMDMnDnDM
DnDgRSMPDnSnDDMgRSnSMDnSnDMgRSnDMDnDMDMgRSRSnDMDMPDMnSnDMnD
nSSnDMgRSMDnSnDgMnDMgMgMnSMPDnDnDnDnDSMgMnDnDnDMgMgRSMPDMSnS
SDMgMgMgMMDnSDnDnDDnDnDgRSRSnDMPDSnDMDnSnDMgRSnDnDRSRSRSRSnD
MDRSDSDnSnDMDSnSRSMggRSSnSSDMDMDMDMnDSSMDSRSnDMDSRSnSDnDMnD
DMPDSRSnDMgRSSnSnDMgMg
MDDnDnDMgMDMPDnSSSDMgMPDMMPDSDSnSMDnDnSnDnSnDnDMgRSSMggRSSM
gMPDSSRSDMDnDnDMPDDMSRSMDMgMDSnSSnDnSnSnDnDSMDnDMDnDnDnDMDS
SnDMDnDMgRSMDMDSRSnSSDMgMPDSnSnSDMgMPDnDnSnDMSDnDMnDnDSRSRSS
MDnDDnDnDgRSDnDMPDMDMPDDnSDSDSnDDMPDnDSnDMDSnDSDnDnSSnDgMDS
DSRSnSDnSnSnSMDMDgMDSMDMgRSRSMDMgRSMDnDnDRSSRSnDnDnDMDnDMgM
nDnSMPDMDMDSDMDMgMgRSnSDMPDMSnDnDSSnSMDDnDMnDSMPDMgRSnDDnD
nDnSRSMnDMMnDMDgRSnDMDMDMgRSRSMnDMgRSDgMgMPDSRSMDDnDnDnSnSn
SnSnDMnDMPDnDSnDMPDDSnDMDn

**Example 2.**
SRSRSDDnDMgMPDMnSnDRSnSMMgMgMPDMPDMDMgRSMDSSnDMDMDMnDnSDS
MnSnSRSMDSRSDMnDMPDMDMPDSnDMMDnDSnDMDMDnDnSMgRSnDMgMDMDM
SMPDMgRSSnDMDSnSRSnSSRSMSRSDMgRSnDMgRSnDMnSRSMDgRSMDnDMgRSnD
MgMgMgMDnSnDMnDMDSRSMDMDMDMDnSnDMDRSMPDMgRSRSMPDMDMDSnD
MgRSnSMgMDgMDMDnDMnDnDSnDDSnSMgMDSMPDnDSnSMDMnSnDSSSRSDSnDS
DgRSDnDSRSnSDnDnSMDMMgMDSnDSSDMPDSRSSnSRSRSnDMgRSSnSnDMDMMgM
DMMPDSMDnDnDMDSnDMDSMgMnDnSMgggRSnSRSMPDnDnDnDMgRSMDMSRSnSD
nnSnSRSSMgRSnDSMDSRSSnDMnSnSnDMgRSMDMDMPDSMPDMgMgMDDSnDSnDnS
MPDnDnDDMPDMggRSnDMMPDMDnDMPDMSnSMPDMPDSSRSSnDMDnDMPDnDnD
MggMgRSnSSnDMgRSDSMDgRSnDMDMDnSnSDnS
RSnSMDnDnnDnDSnDSRSDMgMnDSnDSDSMgMPDMMnDnSDSRSRSDMnDgMgRSMnS
SRSnSnSMDnSMgRSMDnSnSDnDMDRSnDMDMDMgMgMgMDRSSMgRSnDnDMMSRS
DSnSnDMnDMPDMDSnDMgMSMDMgRSRSRSSRSRSRSRSnDMgRSSDMDMgMgMnDn
DnDMnDMgMDnSRSnSnDnDMDMgMDMgRSSnDMMPDDMDMgMgRSnSnDMDnDnSSD
MnDSnDnDMgMggRSnSRSDgMPDSDMDnDSDMDnDnDgRSnDMDSSnDnSnSMgMPDMD
nSnDnDMgRSnDSnDnDMnDnDSRSMnDSnDMgggMDMDDnDMPDSMgMDgMPDSDMDn
nDMPDMDMSnDnDMDSnSnSSMgRSMDMDMDMgRSnSSDMgMgRSRSSRSnDnSnDMDn
DnDSMgRSnDnSnSMDMgMnDSnSnDSn

Similarly ***applying SNCA2*** on the set of 1000 independent $U[0; 1]$ variates, using Sa (S) as the additional starting note, we immediately have a sequence of 1000 notes as our first illustrative example of raga *Bageshree* as given below:

**Example 1.**
SMgRSDnSDnSRSMgRSMgMgRSDnnDSDnnDMgRSDnnDSMgRSSMgMMgMgRSSnnDSg
MMDDnSDnSMSnnDgMDSDnnDMDDnnDSDnSSMgRSDnnDMDDnSMgMgMgMgRSDnnD
SDnSDnSDnSDnSDnSMSnnDMDDnnDSDnSDnnDMDDnnDMDDnSMgMDDnnDMPDMDD
nnDSDnSRnSDnnDMDDnSDnnDSDnnDMDSDnnDMgRSRnDMDDnSDnnDMDDnnDMPDg
MDMDDnnDSSMgRSMgRSDnnDMgMgRSRSSnSMMDDnnDnSMMgRSSMgRSMgMMDD
nnDnnDMDDnnDSDnSMgMDDnnDSMgMgMDDnSMgMDMDDnnDMgRSDnnDSSnnDMP
DgMDDnSRSDnnDSMgRSDnSRnSDnSMgMDSDnnDSDnSRSDnnDnSDnSRnSDnSDnnDSM
gMgMDDnnDMgRSMgRSDnnDSDnSDnnDRnSRSDnnDMgRSSRSDnnDMDDnnDMgRSMg
MgMgRSDnnDMDSDnnDMgRSDnSDnnDSDnnDnnDMgRSDnnDMDDnSRSSRSDnSRDnnD
MDDnSMgMMDSMgRSRSRSSRSMgMgRSDnnDSDnnDSDnnDMPDgMDSDnSDnSMgRS
MMgMgMDMPDMPDggMDDnSRnSMgMDSSRSRSMgMgMDDnSDnSSnnDSDnnDMDDnn
DMDDnnDMPDMDDnSDnSRnSDnnDggMgRSDnSDnnDSSnnDMDSDnnDSMgRSDnnDMD
DnnDMDDnSMgMgMDDnSMgMgMDDnSDnnDMgRSDnSMgRSSnnDSDnnDnnDMgRSMg
RSRSMgRSSRnSRnSRnSDnnDMPDgMgMDMDDnSDnSMgMgMgRSRSSnSRSRnSMgRSD
nnDSDnSRSSRnSDnSDnSMSnnDSDnnDMDMDMPDMPDMDDnnDggMMgMDDnSSRnSD
nnDnSRnSDnnDMDDnnDMDSDnSDnSDnSDnSRnSRnSDnSMgMgRSDnSDnSDnSMgMgRS
RnDgMgRSRnSDnSMgMDDnnDRnSDnSDn

**Example 2.**
gRSDnnDSDnnDMDMDSDnnDMDDnnDMDDnnDSSnnDnnDSMgRSSRnSDnnDMDDnSDn
SMgRSMgMgRSSnnDMDDnSDnnDSSnSDnSDnSMgRSDnnDMgMgRSDnnDMgMDDnSDn
SDnnDMDDnnDMPDMgRSMgRSSRSDnnDMDMDDnSRnSRSMgMDSSMgMMDDnnDMD
DnSMgRSDnnDMDDnSDnnDMDSSnnDRnSDnSDnnDMPDMDDnSMgMMgMDSSMgRSS
MgRSRSDnSDnnDnSDnnDnnDMDDnnDSDnnDSDnSDnnDMDDnnDSDnnDMDSSRSMgMg
RSSnnDMDDnSMgRSMgMgRSDnnDMgMgMDDnnDMDDnSMgMDDnSDnSDnnDMDDnn
DSDnSDnSDnnDMPDMPDMDDnSMgMgRSMgRSMSnnDMgRSSnnDnSMgMgMgMDDnnD
nnDMDDnnDSMgRSMgMgMDDnnDSDnSDnnDSDnnDMDDnnDMDSSRnSDnSRnSSMgMg
RSMgMMgMgMDDnSDnnDSDnSRSSnnDnSDnSDnnDSSnnDMDMDDnnDMDDnnDSSRnS
DnnDMDDnSDnSRnSMgRSDnnDMPDggMDDnnDgMDDnSDnnDMDDnnDSSRnSMgRSDn
nDMPDMPDMDSDnnDSDnSDnnDMDDnSMSnnDMDDnnDSDnnDMDSDnSRnSDnnDRnS
MgRSRnDMDDnnDMPDgMgRSDnnDMDDnnDSDnnDMgRSSnnDSDnSDnnDMDDnnDMD
DnSRnSMgRSRnSMgMgMgMgMDDnnDMDDnnDMDMDDnnDSDnnDnnDSSnSDnnDMDD
nnDMDDnnDSDnSRnSRSDnnDMDMDMDDnnDSDnnDMDSMgRSMgRSMMDDnnDMDDn
nDMDDnSMgRSDnnDnnDMPDgMDDnnDRnSMgMDMDDnnDMDMDSDnSRnDMgRSSnn
DMDDnnDnnDMPDMDDnSMgRSMSnnDSDnSDnnDRnSDnSRSDnSDnSDnnDRnSDnnDM
DDnnDMPDMDDnSRSMgMDDnSRnSDnSDnSDnSMgM

## 6. Observed Characteristics of the Markov Chain

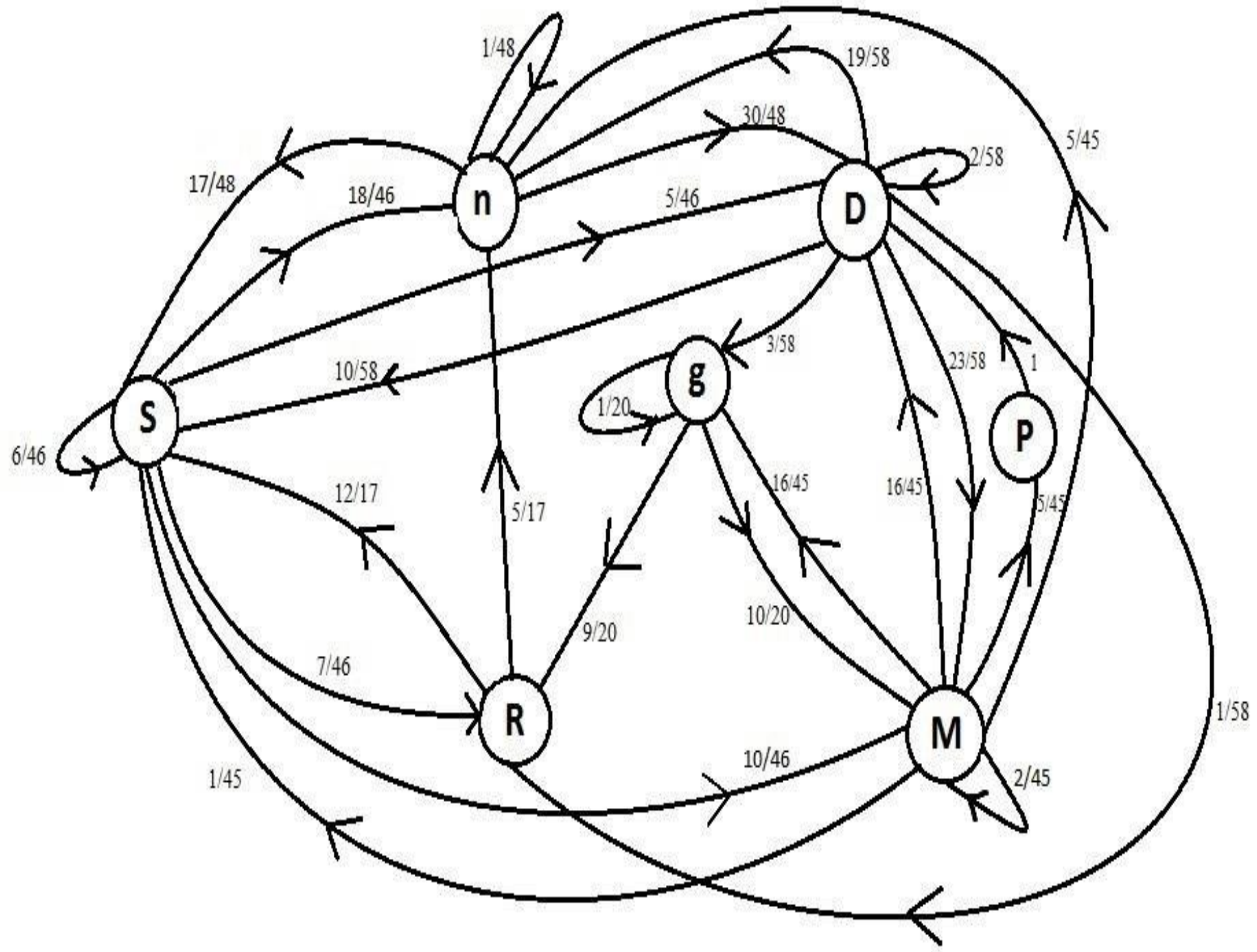

Fig.1. First order Markov State Transition Diagram of Raga Bageshree

***The chain is Ergodic markov chain.***

A Markov chain is called an *ergodic* chain if it is possible to go from every state to every state (not necessarily in one move). Tpm of First order can be inspected and checked that it is possible to go from every state to other state not necessarily in one step. For Ex. It is evident from Fig. 1. There is no direct path from State P to State S,R,g,M,P,n but we can reach to any of these states with 2 step transition i.e. start from reach D ( with 100% ) probability then from D we can reach to any state S,R,g,M,P,n.

***The chain is regular***

A Markov chain is called a *regular* chain if some power of the transition matrix has only positive elements i.e. a square matrix $A$ is called **regular** if for some integer $n$ all entries of $A^n$ are positive. In other words, for some *n*, it is possible to go from any state to any state in exactly *n* steps. *It is clear from this definition that every regular chain is ergodic. On the other hand, an ergodic chain is not necessarily regular*.

**Theorem:** Let $A$ be the transition matrix for a regular chain. Then, as *n tends to infinity*, the powers $A^n$ approach a limiting matrix **W** with all rows the same vector **w**. The vector **w** is a strictly positive probability vector (i.e., the components are all positive and they sum to one).

We created a program to find the nth power of matrix i.e. $A^n$ (A is a tpm of Raga Bageshree with first order markov, TABLE 1). Program was executed till the limiting matrix **W** was not obtained. We found that $28^{th}$ power of the initial tpm matrix (first order) gave the limiting matrix W, with fixed probability vector
**w = [0.192469, 0.071130, 0.083682, 0.188285, 0.242678, 0.200837].**
Starting from any position, after 27 steps, we reach to a steady state condition where transition probability gets fixed as vector **w**.

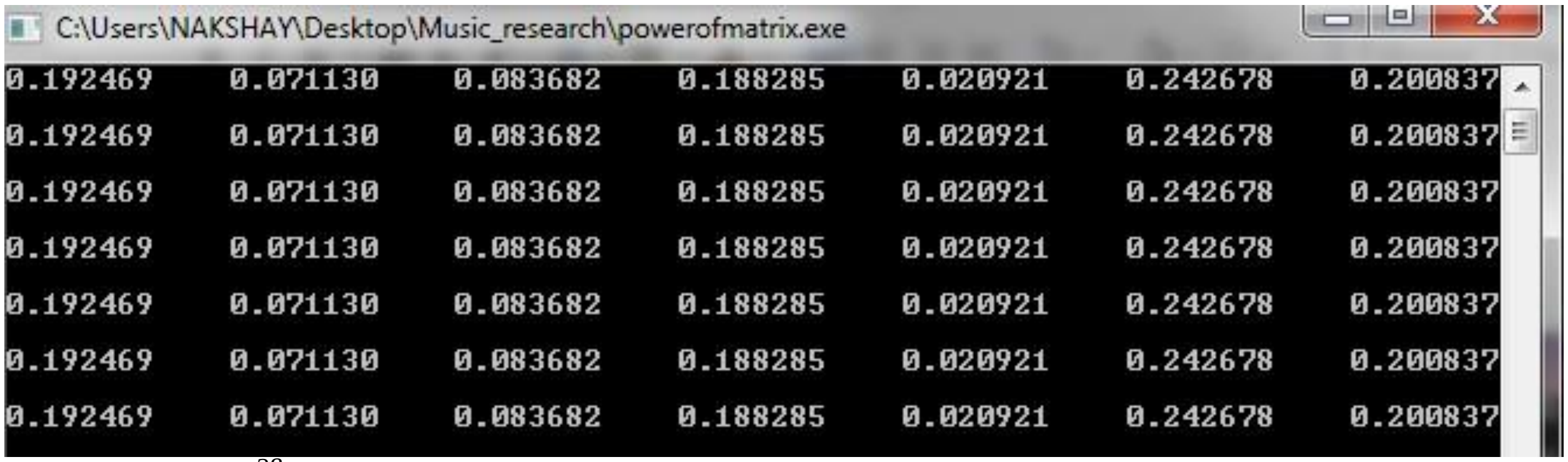
C:\Users\NAKSHAY\Desktop\Music_research\powerofmatrix.exe

| | | | | | | |
|---|---|---|---|---|---|---|
| 0.192469 | 0.071130 | 0.083682 | 0.188285 | 0.020921 | 0.242678 | 0.200837 |
| 0.192469 | 0.071130 | 0.083682 | 0.188285 | 0.020921 | 0.242678 | 0.200837 |
| 0.192469 | 0.071130 | 0.083682 | 0.188285 | 0.020921 | 0.242678 | 0.200837 |
| 0.192469 | 0.071130 | 0.083682 | 0.188285 | 0.020921 | 0.242678 | 0.200837 |
| 0.192469 | 0.071130 | 0.083682 | 0.188285 | 0.020921 | 0.242678 | 0.200837 |
| 0.192469 | 0.071130 | 0.083682 | 0.188285 | 0.020921 | 0.242678 | 0.200837 |
| 0.192469 | 0.071130 | 0.083682 | 0.188285 | 0.020921 | 0.242678 | 0.200837 |

*Screenshot of $A^{28}$ matrix showing fixed probability vector.*

## 7. Concluding Remarks and Supporting Tables

1. There are many statistical tools to verify which order of markov will give the best results, but an order that is statistically valid may not lead to musically meaningful note sequences! Hence, it is left to the composer to test the validity of the generated note sequences by actual hearing. It is quite possible that certain typical note sequences generated by the first order Markov chain may be better but certain other sequences generated by the second order could be handy for the composer. However, increasing the order arbitrarily is not advisable (see point 4 below).

2. SNCA aims at producing the initial note sequences only, i.e. the computer is only giving hint what to play. Musical paraphernalias like duration, loudness, and transition between pitches are completely decided by the composer.

3. The goal of SNCA is not to produce musical compositions which preserves the raga correctly, but to produce raga based melodious compositions (songs).

4. More than 40% transitional probabilities are already zero in tpm of second order markov (table 3), so if we are going to increase the markov order further tpm will take the form of sparse matrix. This would reduce the randomness in the produced sequences, which fails our motive of producing varied raga based note sequences with the help of computer.

5. Novice composers and music trainee face tremendous difficulty in how to begin the composition i.e. the starting musical notes of the song. Semi natural composition can facilitate them in their genuine endeavors of producing good raga based songs. Thus it can be a gigantic step against musical piracy.

6. Singh et. Al (2012) showed that raga based songs have strong therapeutic effects on human mind. SNC (with first and second order markov) will aid this process by giving numerous raga based songs.

7. The Bageshree note sequence is provided in the appendix. This helped in building the tpm.

TABLE 1: BAGESHREE transition probability matrix (Markov First Order)

| | S | R | G | M | P | D | n |
|---|---|---|---|---|---|---|---|
| S | 6/46 | 7/46 | 0/46 | 10/46 | 0/46 | 5/46 | 18/46 |
| R | 12/17 | 0/17 | 0/17 | 0/17 | 0/17 | 0/17 | 5/17 |
| G | 0/20 | 9/20 | 1/20 | 10/20 | 0/20 | 0/20 | 0/20 |
| M | 1/45 | 0/45 | 16/45 | 2/45 | 5/45 | 16/45 | 5/45 |
| P | 0/5 | 0/5 | 0/5 | 0/5 | 0/5 | 5/5 | 0/5 |
| D | 10/58 | 1/58 | 3/58 | 23/58 | 0/58 | 2/58 | 19/58 |
| N | 17/48 | 0/48 | 0/48 | 0/48 | 0/48 | 30/48 | 1/48 |

TABLE 2: Bageshree CLASS matrix (Markov First Order)

| | S | R | G | M | P | D | N |
|---|---|---|---|---|---|---|---|
| S | [0-6/46) | [6/46-13/46) | [13/46-13/46) | [13/46-23/46) | [23/46-23/46) | [23/46-28/46) | [28/46-46/46) |
| R | [0-12/17) | [12/17-12/17) | [12/17-12/17) | [12/17-12/17) | [12/17-12/17) | [12/17-12/17) | [12/17-17/17) |
| G | [0-0/20) | [0-9/20) | [9/20-10/20) | [10/20-20/20) | 0 | 0 | 0 |
| M | [0-1/45) | [1/45-1/45) | [1/45-17/45) | [17/45-19/45) | [19/45-24/45) | [24/45-40/45) | [40/45-45/45) |
| P | 0 | 0 | 0 | 0 | 0 | [0-5/5] | 0 |
| D | [0-10/58) | [10/58-11/58) | [11/58-14/58) | [14/58-37/58) | [37/58-37/58) | [37/58-39/58) | [39/58-58/58) |
| N | [0-17/48) | [17/48-17/48) | [17/48-17/48) | [17/48-17/48) | [17/48-17/48) | [17/48-47/48) | [47/48-48/48) |

TABLE 3: BAGESHREE transition probability matrix (Markov Second Order)

| | S | R | g | M | P | D | n |
|---|---|---|---|---|---|---|---|
| SS | 0/6 | 2/6 | 0/6 | 1/6 | 0/6 | 0/6 | 3/6 |
| SR | 3/7 | 0/7 | 0/7 | 0/7 | 0/7 | 0/7 | 4/7 |
| Sg | 0 | 0 | 0 | 0 | 0 | 0 | 0 |
| SM | 1/10 | 0/10 | 8/10 | 1/10 | 0/10 | 0/10 | 0/10 |
| SP | 0 | 0 | 0 | 0 | 0 | 0 | 0 |
| SD | 0/5 | 0/5 | 0/5 | 0/5 | 0/5 | 0/5 | 5/5 |
| | | | | | | | |
| Sn | 4/18 | 0/18 | 0/18 | 0/18 | 0/18 | 0/18 | 14/18 |
| | | | | | | | |
| RS | 3/12 | 1/12 | 0/12 | 3/12 | 0/12 | 1/12 | 4/12 |
| RR | 0 | 0 | 0 | 0 | 0 | 0 | 0 |
| Rg | 0 | 0 | 0 | 0 | 0 | 0 | 0 |
| RM | 0 | 0 | 0 | 0 | 0 | 0 | 0 |
| RP | 0 | 0 | 0 | 0 | 0 | 0 | 0 |
| RD | 0 | 0 | 0 | 0 | 0 | 0 | 0 |
| Rn | 4/5 | 0/5 | 0/5 | 0/5 | 0/5 | 1/5 | 0/5 |
| | | | | | | | |
| gS | 0 | 0 | 0 | 0 | 0 | 0 | 0 gR |
| 9/9 | 0/9 | 0/9 | 0/9 | 0/9 | 0/9 | 0/9 | |
| Gg | 0/1 | 0/1 | 0/1 | 1/1 | 0/1 | 0/1 | 0/1 |
| gM | 0/10 | 0/10 | 4/10 | 1/10 | 0/10 | 4/10 | 1/10 |
| gP | 0 | 0 | 0 | 0 | 0 | 0 | 0 |
| gD | 0 | 0 | 0 | 0 | 0 | 0 | 0 |
| Gn | 0 | 0 | 0 | 0 | 0 | 0 | 0 |
| | | | | | | | |
| MS | 0/1 | 0/1 | 0/1 | 0/1 | 0/1 | 0/1 | 1/1 |
| MR | 0 | 0 | 0 | 0 | 0 | 0 | 0 |
| Mg | 0/16 | 9/16 | 0/16 | 7/16 | 0/16 | 0/16 | 0/16 |
| MM | 0/2 | 0/2 | ½ | 0/2 | 0/2 | 1/2 | 0/2 |
| MP | 0/5 | 0/5 | 0/5 | 0/5 | 0/5 | 5/5 | 0/5 |
| MD | 2/16 | 0/16 | 0/16 | 2/16 | 0/16 | 2/16 | 10/16 |
| Mn | 0/5 | 0/5 | 0/5 | 0/5 | 0/5 | 4/5 | 1/5 |
| | | | | | | | |
| PS | 0 | 0 | 0 | 0 | 0 | 0 | 0 |
| PR | 0 | 0 | 0 | 0 | 0 | 0 | 0 |
| Pg | 0 | 0 | 0 | 0 | 0 | 0 | 0 |
| PM | 0 | 0 | 0 | 0 | 0 | 0 | 0 |
| PP | 0 | 0 | 0 | 0 | 0 | 0 | 0 |
| PD | 0/5 | 0/5 | 2/5 | 3/5 | 0/5 | 0/5 | 0/5 |
| Pn | 0 | 0 | 0 | 0 | 0 | 0 | 0 |
| | | | | | | | |
| DS | 2/9 | 0/9 | 0/9 | 1/9 | 0/9 | 2/9 | 4/9 |
| DR | 0/1 | 0/1 | 0/1 | 0/1 | 0/1 | 0/1 | 1/1 |
| Dg | 0/3 | 0/3 | 1/3 | 2/3 | 0/3 | 0/3 | 0/3 |
| DM | 0/23 | 0/23 | 3/23 | 0/23 | 5/23 | 11/23 | 4/23 |
| DP | 0 | 0 | 0 | 0 | 0 | 0 | 0 |
| DD | 0/2 | 0/2 | 0/2 | 0/2 | 0/2 | 0/2 | 2/2 |
| Dn | 9/19 | 0/19 | 0/19 | 0/19 | 0/19 | 10/19 | 0/19 |
| | | | | | | | |
| nS | 1/17 | 4/17 | 0/17 | 5/17 | 0/17 | 2/17 | 5/17 |
| nR | 0 | 0 | 0 | 0 | 0 | 0 | 0 |
| ng | 0 | 0 | 0 | 0 | 0 | 0 | 0 |
| nM | 0 | 0 | 0 | 0 | 0 | 0 | 0 |
| nP | 0 | 0 | 0 | 0 | 0 | 0 | 0 |
| nD | 8/30 | 1/30 | 1/30 | 18/30 | 0/30 | 0/30 | 2/30 |
| nn | 0/1 | 0/1 | 0/1 | 0/1 | 0/1 | 1/1 | 0/1 |
| | | | | | | | |

TABLE 4: Bageshree CLASS matrix (Markov Second Order)

| | S | R | G | M | P | D | n |
|---|---|---|---|---|---|---|---|
| SS | 0 | [0-2/6) | [2/6-2/6) | [ 2/6-3/6 ) | [3/6-3/6) | [3/6-3/6) | [3/6-3/6) |
| SR | [0-3/7) | [3/7-3/7) | [3/7-3/7) | [3/7-3/7) | [3/7-3/7) | [3/7-3/7) | [3/7-7/7) |
| Sg | 0 | 0 | 0 | 0 | 0 | 0 | 0 |
| SM | [0-1/10) | [1/10-1/10) | [1/10-9/10) | [9/10-10/10) | 0 | 0 | 0 |
| SP | 0 | 0 | 0 | 0 | 0 | 0 | 0 |
| SD | 0 | 0 | 0 | 0 | 0 | 0 | [0-5/5) |
| Sn | [0-4/18) | [4/18-4/18) | [4/18-4/18) | [4/18-4/18) | [4/18-4/18) | [4/18-4/18) | [4/18-14/18) |
| | | | | | | | |
| RS | [0-3/12) | [3/12-4/12) | [4/12-4/12) | [4/12-7/12) | [7/12-7/12) | [7/12-8/12) | [8/12-12/12) |
| RR | 0 | 0 | 0 | 0 | 0 | 0 | 0 |
| Rg | 0 | 0 | 0 | 0 | 0 | 0 | 0 |
| RM | 0 | 0 | 0 | 0 | 0 | 0 | 0 |
| RP | 0 | 0 | 0 | 0 | 0 | 0 | 0 |
| RD | 0 | 0 | 0 | 0 | 0 | 0 | 0 |
| Rn | [0-4/5) | [4/5-4/5) | [4/5-4/5) | [4/5-4/5) | [4/5-4/5) | [4/5-5/5) | 0 |
| | | | | | | | |
| gS | 0 | 0 | 0 | 0 | 0 | 0 | 0 |
| gR | [0-9/9) | 0 | 0 | 0 | 0 | 0 | 0 |
| gg | 0 | 0 | 0 | [0-1) | 0 | 0 | 0 |
| gM | 0 | 0 | [0-4/10) | [4/10-5/10) | [5/10-5/10) | [5/10-9/10) | [9/10-10/10) |
| gP | 0 | 0 | 0 | 0 | 0 | 0 | 0 |
| gD | 0 | 0 | 0 | 0 | 0 | 0 | 0 |
| gn | 0 | 0 | 0 | 0 | 0 | 0 | 0 |
| | | | | | | | |
| MS | 0 | 0 | 0 | 0 | 0 | 0 | [0-1) |
| MR | 0 | 0 | 0 | 0 | 0 | 0 | 0 |
| Mg | 0 | [0-9/16) | [9/16- | [9/16- | 0 | 0 | 0 |
| | | | 9/16) | 16/16) | | | |
| MM | 0 | 0 | [0-1/2) | [1/2-1/2) | [1/2-1/2) | [1/2-1) | 0 |
| MP | 0 | 0 | 0 | 0 | 0 | [0-5/5) | 0 |
| MD | [0-2/16) | [2/16-2/16) | [2/16-2/16) | [2/16-4/16) | [4/16-4/16) | [4/16-6/16) | [6/16-16/16) |
| Mn | 0 | 0 | 0 | 0 | 0 | [0-4/5) | [4/5-5/5) |
| | | | | | | | |
| PS | 0 | 0 | 0 | 0 | 0 | 0 | 0 |
| PR | 0 | 0 | 0 | 0 | 0 | 0 | 0 |
| Pg | 0 | 0 | 0 | 0 | 0 | 0 | 0 |
| PM | 0 | 0 | 0 | 0 | 0 | 0 | 0 |
| PP | 0 | 0 | 0 | 0 | 0 | 0 | 0 |
| PD | 0 | 0 | [0-2/5) | [2/5-5/5) | 0 | 0 | 0 |
| Pn | 0 | 0 | 0 | 0 | 0 | 0 | 0 |
| | | | | | | | |
| DS | [0-2/9) | [2/9-2/9) | [2/9-2/9) | [) | [) | [) | [) |
| DR | 0 | 0 | 0 | 0 | 0 | 0 | [0-1/1) |
| Dg | 0 | 0 | [0-1/3) | [1/3-3/3) | 0 | 0 | 0 |
| DM | 0 | 0 | [0-3/23) | [3/23-3/23) | [3/23-8/23) | [8/23-19/23) | [19/23-23/23) |
| DP | 0 | 0 | 0 | 0 | 0 | 0 | 0 |
| DD | 0 | 0 | 0 | 0 | 0 | 0 | [0-2/2) |
| Dn | [0-9/19) | [9/19-9/19) | [9/19-9/19) | [9/19-9/19) | [9/19-9/19) | [9/19-19/19) | 0 |
| | | | | | | | |
| nS | [0-1/17) | [1/17-5/17) | [5/17-5/17) | [5/17-10/17) | [10/17-10/17) | [10/17-12/17) | [12/17-17/17) |
| nR | 0 | 0 | 0 | 0 | 0 | 0 | 0 |
| ng | 0 | 0 | 0 | 0 | 0 | 0 | 0 |
| nM | 0 | 0 | 0 | 0 | 0 | 0 | 0 |
| nP | 0 | 0 | 0 | 0 | 0 | 0 | 0 |
| nD | [0-8/30) | [8/30-9/30) | [9/30-10/30) | [10/30-28/30) | [28/30-28/30) | [28/30-28/30) | [28/30-30/30) |
| nn | 0 | 0 | 0 | 0 | 0 | [0-1/1) | 0 |
| | | | | | | | |

Appendix: Bageshree note sequence (Dutta, 2006)

| Instance | Note | Pitch | Instance | Note | Pitch | Instance | Note | Pitch | Instance | Note | Pitch |
|---|---|---|---|---|---|---|---|---|---|---|---|
| (t) | | ($Y_t$) | (t) | | ($Y_t$) | (t) | | ($Y_t$) | (t) | | ($Y_t$) |
| 1 | S | 0 | 61 | S | 0 | 121 | **R** | **14** | 181 | **S** | **12** |
| 2 | *n* | *-2* | 62 | *n* | -2 | 122 | n | 10 | 182 | n | 10 |
| 3 | *D* | *-3* | 63 | *D* | -3 | 123 | **S** | **12** | 183 | D | 9 |
| 4 | *n* | *-2* | 64 | S | 0 | 124 | n | 10 | 184 | **R** | **14** |
| 5 | S | 0 | 65 | *n* | -2 | 125 | D | 9 | 185 | n | 10 |
| 6 | M | 5 | 66 | S | 0 | 126 | M | 5 | 186 | D | 9 |
| 7 | M | 5 | 67 | M | 5 | 127 | D | 9 | 187 | M | 5 |
| 8 | g | 3 | 68 | g | 3 | 128 | n | 10 | 188 | D | 9 |
| 9 | M | 5 | 69 | M | 5 | 129 | D | 9 | 189 | M | 5 |
| 10 | D | 9 | 70 | D | 9 | 130 | g | 3 | 190 | n | 10 |
| 11 | n | 10 | 71 | n | 10 | 131 | g | 3 | 191 | n | 10 |
| 12 | D | 9 | 72 | D | 9 | 132 | M | 5 | 192 | D | 9 |
| 13 | M | 5 | 73 | M | 5 | 133 | M | 5 | 193 | M | 5 |
| 14 | n | 10 | 74 | n | 10 | 134 | D | 9 | 194 | D | 9 |
| 15 | D | 9 | 75 | D | 9 | 135 | D | 9 | 195 | **S** | **12** |
| 16 | M | 5 | 76 | M | 5 | 136 | n | 10 | 196 | **S** | **12** |
| 17 | D | 9 | 77 | P | 7 | 137 | D | 9 | 197 | **M** | **17** |
| 18 | n | 10 | 78 | D | 9 | 138 | **S** | **12** | 198 | **g** | **15** |
| 19 | **S** | **12** | 79 | g | 3 | 139 | D | 9 | 199 | **M** | **17** |
| 20 | **S** | **12** | 80 | M | 5 | 140 | n | 10 | 200 | **g** | **15** |
| 21 | n | 10 | 81 | g | 3 | 141 | **S** | **12** | 201 | **R** | **14** |
| 22 | D | 9 | 82 | R | 2 | 142 | D | 9 | 202 | **S** | **12** |
| 23 | M | 5 | 83 | S | 0 | 143 | n | 10 | 203 | **S** | **12** |
| 24 | P | 7 | 84 | S | 0 | 144 | D | 9 | 204 | **R** | **14** |
| 25 | D | 9 | 85 | R | 2 | 145 | **S** | **12** | 205 | n | 10 |
| 26 | M | 5 | 86 | S | 0 | 146 | n | 10 | 206 | **S** | **12** |
| 27 | g | 3 | 87 | *n* | -2 | 147 | D | 9 | 207 | n | 10 |
| 28 | M | 5 | 88 | *D* | -3 | 148 | M | 5 | 208 | D | 9 |
| 29 | g | 3 | 89 | *M* | -7 | 149 | D | 9 | 209 | **S** | **12** |
| | | | | | | | | | | | |
| 30 | R | 2 | 90 | *D* | -3 | 150 | n | 10 | 210 | n | 10 |
| 31 | S | 0 | 91 | *n* | -2 | 151 | **S** | **12** | 211 | D | 9 |
| 32 | *D* | *-3* | 92 | *D* | -3 | 152 | n | 10 | 212 | M | 5 |
| 33 | *n* | *-2* | 93 | S | 0 | 153 | D | 9 | 213 | D | 9 |
| 34 | S | 0 | 94 | M | 5 | 154 | M | 5 | 214 | n | 10 |
| 35 | M | 5 | 95 | g | 3 | 155 | P | 7 | 215 | D | 9 |
| 36 | S | 0 | 96 | M | 5 | 156 | D | 9 | 216 | **S** | **12** |
| 37 | *n* | *-2* | 97 | D | 9 | 157 | M | 5 | 217 | D | 9 |
| 38 | *D* | *-3* | 98 | M | 5 | 158 | g | 3 | 218 | n | 10 |
| 39 | *M* | *-7* | 99 | D | 9 | 159 | R | 2 | 219 | **S** | **12** |
| 40 | *D* | *-3* | 100 | D | 9 | 160 | S | 0 | 220 | D | 9 |
| 41 | *n* | *-2* | 101 | n | 10 | 161 | M | 5 | 221 | n | 10 |
| 42 | *D* | -3 | 102 | D | 9 | 162 | g | 3 | 222 | D | 9 |
| 43 | *M* | -7 | 103 | M | 5 | 163 | M | 5 | 223 | **S** | **12** |
| 44 | *n* | -2 | 104 | P | 7 | 164 | D | 9 | 224 | n | 10 |
| 45 | *D* | -3 | 105 | D | 9 | 165 | n | 10 | 225 | **S** | **12** |
| 46 | *M* | -7 | 106 | g | 3 | 166 | **S** | **12** | 226 | **R** | **14** |
| 47 | *D* | -3 | 107 | M | 5 | 167 | **R** | **14** | 227 | n | 10 |
| 48 | *n* | -2 | 108 | g | 3 | 168 | **S** | **12** | 228 | **S** | **12** |
| 49 | S | 0 | 109 | R | 2 | 169 | **M** | **17** | 229 | n | 10 |
| 50 | M | 5 | 110 | S | 0 | 170 | **g** | **15** | 230 | D | 9 |
| 51 | g | 3 | 111 | S | 0 | 171 | **R** | **14** | 231 | M | 5 |
| 52 | R | 2 | 112 | *n* | -2 | 172 | **S** | **12** | 232 | P | 7 |
| 53 | S | 0 | 113 | S | 0 | 173 | **M** | **17** | 233 | D | 9 |
| 54 | R | 2 | 114 | M | 5 | 174 | **g** | **15** | 234 | M | 5 |
| 55 | S | 0 | 115 | g | 3 | 175 | **R** | **14** | 235 | g | 3 |
| 56 | *n* | -2 | 116 | M | 5 | 176 | **S** | **12** | 236 | R | 2 |
| 57 | *D* | -3 | 117 | n | 10 | 177 | n | 10 | 237 | S | 0 |
| 58 | *M* | -7 | 118 | D | 9 | 178 | **S** | **12** | 238 | *n* | -2 |
| 59 | *D* | -3 | 119 | n | 10 | 179 | **R** | **14** | 239 | *D* | -3 |
| 60 | S | 0 | 120 | **S** | **12** | 180 | n | 10 | 240 | S | 0 |
| | | | | | | | | | | | |